\documentclass[prb, twocolumn,showpacs,preprintnumbers,amsmath,amssymb ]{revtex4}

\usepackage{graphicx}
\usepackage{dcolumn}
\usepackage{bm}

\begin{document}

\input epsf.sty



\title{Long-range potential fluctuations and $1/f$ noise in 
hydrogenated amorphous silicon}

\author{B. V. Fine$^{1,3}$\cite{e.mail}, 
J. P. R. Bakker$^{2}$, and J. I. Dijkhuis$^{2}$}

\affiliation{
$^{1}$Spinoza Institute and $^{2}$Debye Institute, 
Utrecht University, P.O. Box 80000, 3508 TA Utrecht, 
Netherlands \\ 
\mbox{$^{3}$Max Plank Institute for the Physics of Complex Systems,
Noethnitzer Str. 38, D-01187, Dresden, Germany}}

\date{June 4, 2003}

\begin{abstract}
We present a microscopic theory
of the low-frequency voltage noise (known as ``$1/f$'' noise) 
in $\mu$m-thick
films of hydrogenated amorphous silicon. 
This theory  traces the noise back to the long-range fluctuations
of the Coulomb potential created by deep defects, thereby
predicting the {\it absolute} noise intensity  as a function of 
the distribution of defect
activation energies. 
The predictions of this theory are in very good
agreement with our own experiments in terms of both the absolute
intensity and the temperature dependence of the noise spectra.
\end{abstract}
\pacs{71.55.Jv, 72.70.+m, 73.50.Td, 73.61.Jc}


\maketitle


\section{introduction}

When electric current flows through a resistor,
strong low-frequency voltage noise
superposed on the thermal Johnson-Nyquist noise\cite{Johnson,Nyquist} 
is generally observed.
The spectrum  of that excess noise has shape close to
$1/f$, where $f$ is frequency\cite{Weissman}.
In recent decades, the model, according to which
the $1/f$ noise is produced by an ensemble of two-state systems 
having broadly  distributed activation energies 
(BDAE)\cite{vanderZiel,duPre,DDH}, 
has enjoyed a lot of success.
However, the BDAE model  addresses neither the origin 
of the two-state systems nor the noise  mechanism, 
thus leaving
the absolute noise intensity as
an adjustable parameter. The lack 
of direct microscopic calculations predicting
the absolute noise intensity, while being a glaring theoretical gap as such,
also fuels an old but still continuing debate over the question of 
what actually
fluctuates --- the number of carriers
or their mobility\cite{Weissman}. 

In this work, we present a theoretical and experimental investigation
of  $1/f$ noise  in hydrogenated
amorphous silicon  \mbox{(a-Si:H)}. 
Our theory links the noise to the 
fluctuations of the number of carriers, 
predicts the absolute noise intensity, and also allows us to 
extract novel and useful information about the defects 
in this important material\cite{Street}. 

This theory  can be outlined as follows:
Hydrogenated amorphous silicon has a significant number
of deep defects known as dangling bonds.
Thermal fluctuations of the numbers of electrons occupying 
the dangling bonds cause 
long-range potential fluctuations, which 
give rise to  fluctuations of the local densities of carriers, 
which, in turn, lead to resistance fluctuations, which, 
in the presence of current, manifest themselves as voltage noise.
It is the inclusion of the long-range potential fluctuations
into the above scheme that distinguishes our treatment from
many similar theoretical proposals.

\section{Experimental setup}

The $1/f$ noise in a-Si:H has been studied in the literature in a variety of
experimental settings
(see e.g. Refs.~\cite{BA,BAV,KK,Verleg,GJK,Johanson,Goennenwein,BK,Kasap}), 
and exhibited certain features, which depend on numerous details
of each particular experiment. 
In this work we present a fully developed quantitative
study of only one situation, 
which corresponds to our own experiments. Other experimental settings will 
only be 
discussed briefly in the end of the paper.

The present study is focused on an $n-i-n$ 
film of \mbox{a-Si:H}, where $n$ stands for a  40nm-thick 
electron 
doped layer, and $i$  for an  undoped layer of thickness
\mbox{$d = 0.91\mu$m}. The $n-i-n$ structure
is grown by plasma enhanced chemical vapor deposition (PECVD) 
on a highly conductive wafer of crystaline silicon. 
The contact layer
on the top of the structure consists of a 30nm-thick film of titanium
followed by 30nm-thick film of copper.  
The film has area $A=0.56\hbox{cm}^2$. 
It is thermally annealed and then protected from light. 
We observe and analyse the voltage noise spectra 
at frequencies $f = 1 \div 10^4\,$Hz and temperatures
$T = 340 \div 434\,$K  in the presence of electric current 
flowing perpendicular to the plane of the film.
Other details of our experimental setup are described in Ref.~\cite{Verleg}.

\section{Formulation of the theoretical problem}

Now we turn to the theoretical derivation of the noise spectrum
under the above experimental conditions. The central and, presumably,
quite general  part of this derivation is Section~\ref{cf}. 
Most of the rest is specific to
a-Si:H and to the experimental setting considered.

Our goal is to compute  voltage noise spectrum $S_V(f)$ expressed as:
\begin{equation}
{S_V(f) \over V^2} = 4 \int_0^\infty C_V(t) \hbox{cos}(2 \pi f t) \ dt,
\label{Sv}
\end{equation}
where $V$ is the applied voltage, and 
\begin{equation}
C_V (t) = { \langle \delta R(t) \delta R(0) \rangle \over R_0^2}.
\label{Cv1}
\end{equation}
Here, 
$R_0$ is the average resistance of the film, and $\delta R(t)$ is the 
equilibrium resistance noise. As usual\cite{Weissman}, 
the link between the resistance noise
and the voltage noise is  established experimentally by observing that 
$S_V(f) \propto V^2$. 

\begin{figure}
\setlength{\unitlength}{0.1cm}
\begin{picture}(50, 49)
{
\put(-21, -3){
\epsfxsize= 3.5in
\epsfbox{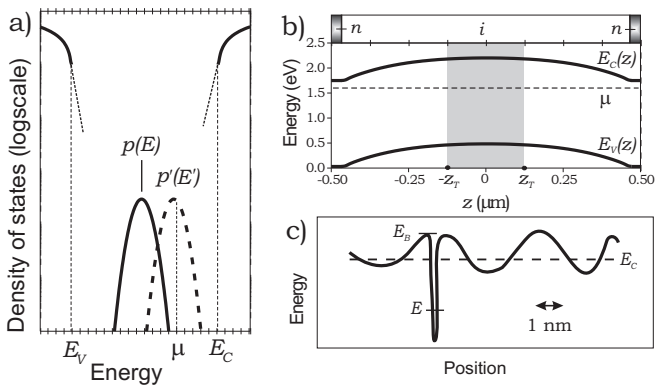}
 }
}

\end{picture} 
\caption{(a) Sketch of the 
density of states in the center of the undoped layer.
(b) Band bending profile. The gray stripe 
represents the ``uniform resistivity layer'' defined in the text.
(c) Cartoon of a deep defect surrounded by medium-range structural
disorder. Note: $e \phi(t, {\mathbf{r}})$ fluctuates on a much longer
lengthscale and with much smaller amplitude.} 
\label{fig1} 
\end{figure}

The density of states of undoped a-Si:H is shown in Fig.~\ref{fig1}a.
It is characterized by a band gap of $1.8\,$~eV 
between the mobility edges $E_v$ and $E_c$
in the valence and conduction bands, respectively. 
The proximity of the $n$-layers induces  
band bending in the 
undoped layer as shown in Fig~\ref{fig1}b.
Around the center
of the film $E_c(z) = E_{c0} - \beta z^2$, 
where 
the $z$-axis is directed perpendicular to the film,
$E_{c0} = \mu + 0.63\,$eV, $\mu$ is the chemical potential,
and $\beta = 1.6\,$eV/$\mu$m$^2$.
The difference $E_{c0} - \mu$ was obtained
experimentally as the conductivity activation energy,
while the value of $\beta$ was found numerically (cf.\cite{BHD}).

Since $\mu$ is significantly closer to $E_c$ than to $E_v$, 
the resistivity $\rho$ is inversely proportional to 
the density of conduction electrons $n_e$.
Keeping in mind that 
\linebreak[4] 
\mbox{$n_e \propto \hbox{exp}\left( - {E_c - \mu \over k_B T} \right)$}, 
we obtain
$\rho(z) = X \hbox{exp} \left( 
{E_c(z)  - \mu \over k_B T} \right)$, 
which is strongly peaked around $z=0$.
Here $X$ is a proportionality coefficient, and $k_B$ is 
the Boltzmann constant.
The total film resistance can then be found as 
$ R_0  = {1 \over A} \int \rho(z) dz \approx  
{\rho(0) \over A} \sqrt{{\pi k_B T \over \beta}}$.

To simplify the treatment we replace the 
actual profile of $\rho(z)$
by  a layer
of constant resistivity \mbox{$\rho_0 = \rho(0)$}
spreading along the $z$-axis between $-z_T$ and $z_T$ (see Fig.~\ref{fig1}b).
We require
the resistance of this constant resistivity layer to be equal to 
$R_0$,
which gives
\begin{equation}
z_T = {1\over 2} \sqrt{{\pi k_B T \over \beta}}
\label{zT}
\end{equation}
(typical value: $0.13 \mu$m).

\section{correlation function}
\label{cf}

The resistivity  within the
``constant resistivity layer'' still fluctuates
as a consequence of the  fluctuations  of 
the screened Coulomb potential $\phi(t, {\mathbf{r}})$ 
created by deep defects:
\begin{equation}
\phi(t, {\mathbf{r}}) = \sum_i {\Delta q_i(t) \over 
\epsilon |{\mathbf{r}} - {\mathbf{a}}_i|}
\hbox{exp} \left( -{|{\mathbf{r}} - {\mathbf{a}}_i| \over r_{s_i} }  \right).
\label{phi}
\end{equation}
Here 
$\Delta q_i(t)$ is the fluctuation of the $i$th defect charge
with respect
to its average value, ${\mathbf{a}}_i$  the position   the 
defect, $r_{s_i}$ the screening radius of that defect, and
\mbox{$\epsilon =  12$} the dielectric constant.
The defects causing the potential fluctuations
may be located outside of the constant resistivity layer.
[Note: the above formula  does not include large static
component of the random Coulomb potential.]

When the local potential $\phi(t, {\mathbf{r}})$ 
 fluctuates, 
the mobility  edge
tracks it, i.e.
$E_c(t, {\mathbf{r}}) = E_{c0} + e \phi(t, {\mathbf{r}})$,
where  $e$ is the electron charge.
Since $\mu$ does not shift with $e \phi(t, {\mathbf{r}})$, 
the density of conduction electrons re-equilibrates  
following $E_c(t, {\mathbf{r}})$ on the timescale of electron drift
from the uniform resistivity layer to the $n$-layers (and then to 
the contact layers). 
Because of the strong band bending,
the drift takes less than  $10^{-7}$s,
i.e. the re-equilibration is effectively instantaneous on the 
timescales of the noise studied \mbox{(${2 \pi \over f} \sim 10^{-4}\div 1$~s).} 
The resistivity can thus be rewritten as
$\rho(t, {\mathbf{r}}) = 
X \exp \left({E_c(t, {\mathbf{r}}) - \mu \over k_B T} \right)$.
Assuming for a moment [and proving later] that 
$|e \phi(t, {\mathbf{r}})| \ll k_B T$ , we expand
$\rho(t, {\mathbf{r}}) = \rho_0 + \delta \rho(t, {\mathbf{r}})$,
where 
$\rho_0 = X \hbox{exp} \left( {E_{c0}  - \mu \over k_B T} \right)$,
and
\begin{equation}
\delta \rho(t, {\mathbf{r}}) = 
{e \phi(t, {\mathbf{r}}) \over k_B T} \rho_0.
\label{drho1}
\end{equation}
For $\delta \rho \ll \rho_0$, the fluctuation of the total resistance is
\begin{equation}
\delta R(t) =  {1 \over A^2} 
\int_{\cal{V}} \delta \rho(t, {\mathbf{r}})  d^3 r,
\label{dR}
\end{equation}
where $\cal{V}$ is the space inside
the constant resistivity layer.
Substituting $R_0 = {2 z_T \rho_0 \over A}$ and 
$\delta R(t)$ given by Eq.(\ref{dR}) 
into Eq.(\ref{Cv1}) 
and then using
Eq.(\ref{drho1}), we obtain
\begin{equation}
C_V(t) =  
\left({e
\over
2 k_B T z_T A }\right)^2
\int_{\cal{V}} d^3 {\mathbf{r}} \int_{\cal{V}}  d^3 {\mathbf{r}}^{\prime}  
\langle \phi(t, {\mathbf{r}})  \phi(0, {\mathbf{r}}^{\prime}) \rangle.
\label{Cvphi}
\end{equation}

\section{Description of Defects}
\label{defects}

In order to evaluate 
$\langle \phi(t, {\mathbf{r}})  \phi(0, {\mathbf{r}}^{\prime}) \rangle$,
we have to  describe the deep defects in the undoped
layer of \mbox{a-Si:H\cite{numbers}}. 

There exist both theoretical arguments and experimental evidence\cite{SJT}
indicating that the concentration of defects in the undoped layer
should be greater in the vicinity of the $n$-layers 
and then decay towards the center on the
lengthscale of about $0.5\mu$m.
Here, however, in order to simplify the theoretical treatment,
we assume that the defect concentration across the entire undoped
layer has a constant value, which we estimate as\cite{numbers}
\mbox{$n_D = 6 \times 10^{15}\,\hbox{cm}^{-3}$}.

Each defect  has 4 possible states: one zero-electron state $D^+$,
two one-electron states $D^0$, and 
one two-electron state, $D^-$. 
We assume a Gaussian probability distribution
\begin{equation}
p(E) = {1 \over \sqrt{2 \pi} \Delta E } \  
\hbox{\textsl{e}}^{-{ (E-E_0)^2 \over 2 \Delta E^2}}
\label{pE}
\end{equation}
for energy $E$ of an electron occupying the $D^0$ state.  
Here $E_0 = \mu - 0.22\,$eV and \mbox{$\Delta E = 0.15\,$eV.}
Energy $E$,  as such,  is associated only with
the $D^+ \leftrightarrow D^0$ transition.
The $D^0 \leftrightarrow D^-$ transition, which
requires capturing the second electron  by the same defect, 
is characterized by energy 
\mbox{$E^{\prime} = E + U$},
where \mbox{$U = 0.2\,$eV} is the correlation energy.
Therefore, the corresponding probability distribution is
\begin{equation}
p^{\prime}(E^{\prime}) = {1 \over \sqrt{2 \pi} \Delta E }
\hbox{\textsl{e}}^{-{ (E^\prime-E_0 - U)^2 \over 2 \Delta E^2}}.
\label{ppEp}
\end{equation}
We shall treat the \mbox{$D^+ \leftrightarrow D^0$} transitions
independently from the \mbox{$D^0 \leftrightarrow D^-$}
transitions, which is justified  as long as 
exp$\left(- {U \over k_B T} \right) \ll 1$.

The noise spectrum obtained later will depend only weakly on 
defect parameters
$n_D$, $E_0$, $U$ and $\Delta E$, mainly through a weak dependence 
on one 
combination of them --- the density
of defect states at the chemical potential. 
That combination, in turn, 
is not very sensitive to the choice of $E_0$ and $U$.

Using
the equilibrium description\cite{AY}  of  
the $D^+$, $D^0$ and $D^-$ states, 
we obtain the mean squared charge fluctuations 
for the  $D^+ \leftrightarrow D^0$ transition:
\begin{equation}
\langle \Delta q^2(E) \rangle_+  = 
 {1 \over 4} \, e^2 \,  \hbox{sech}^2  \left({E - \mu \over 2 k_B T}
 - c \right) ;
\label{Dq2E+0}
\end{equation}
where $c =  {1 \over 2}\, \hbox{ln} 2$. For 
the $D^0 \leftrightarrow D^-$ transition:
\begin{equation}
\langle \Delta q^2(E^{\prime} ) \rangle_-  = 
{1 \over 4} \, e^2  \, 
\hbox{sech}^2  \left({E^{\prime} - \mu \over 2 k_B T} + c \right).
\label{Dq2E0-}
\end{equation}

In order to escape from a deep defect, an electron should reach 
the mobility edge $E_c$.  However, the activation barriers $E_B$
(indicated in Fig.\ref{fig1}c) can vary
as a
result of the {\it medium-range} disorder of 
the amorphous structure (on a lengthscale of $1 \div 10\,$nm). 
We  assume a Gaussian probability
distribution for the values of $E_B$:
\begin{equation}
P(E_B) =  
{ 1 \over \sqrt{2 \pi} \Delta E_B} 
\exp \left( -{(E_B - E_{B0})^2 \over 2  \Delta E_B^2} \right).
\label{PEB}
\end{equation}
where $E_{B0}$ and $\Delta E_B$ are 
to be  extracted from the experimental spectra. These are the only
two adjustable parameters in our treatment. 
They will affect the spectral shape, but not the integrated
noise intensity. 

The fluctuation rate for
the $D^+ \leftrightarrow D^0$ transition is\cite{RBZ}
\begin{equation}
{1 \over \tau_+(E, E_B)} = 
\omega_0(E_B)
\left[ \hbox{\textsl{e}}^{- {E_B - \mu \over k_B T} } +
{1 \over 2} \, \hbox{\textsl{e}}^{ - {E_B - E \over k_B T} } \right];
\label{tau+0}
\end{equation}
and for the $D^0 \leftrightarrow D^-$ transition:
\begin{equation}
{1 \over \tau_-(E^{\prime}, E_B)}  =  
\omega_0(E_B)
\left[ {1 \over 2} \,  \hbox{\textsl{e}}^{ - {E_B - \mu \over k_B T} } +
\hbox{\textsl{e}}^{ - {E_B - E^{\prime} \over k_B T} } \right],
\label{tau0-}
\end{equation}
where the attempt frequency is estimated from the detailed balance condition 
as
\begin{equation}
\omega_0(E_B) = v_{th} \ \sigma \ {\cal N}_c(E_B)
\label{om0}
\end{equation}
Here, $v_{th} = \sqrt{3 k_B T/ m_e^*}$ is the thermal velocity of
conduction electrons;
$m_e^*$  
their effective mass 
\mbox{($m_e^* \approx 0.4 m_e = 3.64 \cdot 10^{-28}$g);}
$\sigma = 10^{-15}\,\hbox{cm}^2$ 
the cross-section of electron capture
by $D^+$ or $D^0$ defect; and
${\cal N}_c(E_B) \approx N_c(E_B) \, k_B T$ the ``concentration'' 
of thermally accessible electronic states
with energies above $E_B$;
$N_c(E_B) = N_c(E_{c0}) \sqrt{(E_B -  E_{c0} + \varepsilon_c)/\varepsilon_c}$
the empirical fit to the density of states  above the mobility edge;
$N_c(E_{c0}) = 4 \times 10^{21}\, \hbox{eV}^{-1} \hbox{cm}^{-3}$;
and 
\mbox{$\varepsilon_c =  0.02 \, \hbox{eV}$}.  
The typical value of  $\omega_0(E_B)$ is then $10^{13}\,\hbox{s}^{-1}$.

Next we evaluate the screening radius. 
Since 
the screening by the conduction electrons in the undoped
layer  can be neglected in view of their very
small concentration ($10^{10} \div 10^{13}$cm$^{-3}$),
two other screening mechanisms should be considered,
namely: (i) by the $n$-layers together with the contact layers; 
and (ii)
by the defects in the undoped layer.

We describe the first mechanism as a perfect screening 
by metallic surfaces.
In other words, we assume that
a defect in the undoped layer
is screened by the infinite set of its 
mirror images constructed with respect to the
mirror planes located at $z = \pm d/2$. 
Although the screening law due to this mechanism is not exponential,
we obtain the best possible value
for $r_s$ entering Eq.(\ref{phi})
as the distance from the defect to the point,
where the screened potential is factor of 
$\hbox{\textsl{e}} =2.718...$  smaller
than the bare Coulomb potential. We have found numerically that $r_s$ has 
angular dependence, which is such that, for a defect
located at $z=0$, 
$r_{sx} = r_{sy} = 0.54 d$, and $ r_{sz} = 0.38 d$. 
We then approximate the screening radius \mbox{due to this 
mechanism as}
\begin{equation}
r_{s1} = ( r_{sx} r_{sy} r_{sz} )^{1/3} = 0.48 d.
\label{rs1}
\end{equation}

The second screening mechanism is similar to that of Thomas-Fermi.
It can be described as follows:
When the potential fluctuates  due to  charge fluctuation on 
a given defect,  the
energies $E$ and $E^{\prime}$ of the neighboring defects become shifted.
In response, those neighboring defects change their occupation numbers 
thus screening the potential of the defect, which  
originally caused the fluctuation.
The equilibrium description of this mechanism \cite{SE} 
results in the screening radius
\begin{equation}
r_{_{\hbox{eq}}} = \sqrt{ { \epsilon \over 4 \pi e^2  \ n_D \nu(T) }} ,
\label{rseq}
\end{equation}
where
\begin{equation}
\label{nu}
\nu(T)   =   {1 \over 4 k_B T}   \left\{ 
\int_{- \infty}^{\infty} 
\tilde{p}(E) dE +
\int_{- \infty}^{\infty} 
\tilde{p}^{\prime}(E^{\prime}) dE^{\prime} 
\right\} ; 
\end{equation}
\begin{equation}
\tilde{p}(E) = p(E) \, \hbox{sech}^2 \left({E - \mu \over 2 k_B T}
 - c \right), 
\label{ptildeE}  
\end{equation}
and
\begin{equation}
\tilde{p}^{\prime}(E^{\prime}) = p^{\prime}(E^{\prime}) \,
\hbox{sech}^2 \left({E^{\prime} - \mu \over 2 k_B T} + c \right).
\label{pptildeEp}
\end{equation}  
The value of $\nu(T)$ depends on $T$ very weakly. 
The product $n_D \nu(0)$ should be recognized as the density
of the defect states at the chemical potential.

The equilibrium description of screening is applicable, when
the sources of the Coulomb potential 
are static, and thus all the defects can contribute to screening.
However, if a given defect fluctuates on time scale $\tau$, its 
potential can only be screened by other defects fluctuating 
on timescales not slower than $\tau$. 
To take this observation into account,
we simply multiply $n_D$ 
in formula (\ref{rseq}) by $b(\tau)$, the fraction of
defects that fluctuate fast enough to take part in the screening
of a defect characterized by the fluctuation time $\tau$.
We estimate that fraction as
\begin{eqnarray}
\nonumber
b(\tau)  =  &
{1 \over 4 k_B T \nu(T)}  \left\{ 
\int_{{\cal E}_+(\tau)} \tilde{p}(E) \, P(E_B) \, dE \, dE_B  
\ \ \ \ \ \  \right. \\
& 
\left.   +
\int_{{\cal E}_-(\tau)} \tilde{p}^{\prime}(E^{\prime}) \,  P(E_B) \, dE^{\prime} 
\, dE_B
\right\},
\label{b}
\end{eqnarray}
where ${\cal E}_+(\tau)$ is the integration region
limited by condition $\tau_+(E, E_B) < \tau$, and 
${\cal E}_-(\tau)$  the integration region limited
by $\tau_-(E^{\prime}, E_B) < \tau$. 
The screening radius due to the second mechanism thus becomes
\begin{equation}
r_{s2}(\tau) = \sqrt{{\epsilon \over 4 \pi e^2 n_D \nu(T) b(\tau) }}.
\label{rsd4}
\end{equation} 

For the slower fraction of defects,  $r_{s2}(\tau)$
approaches  $r_{_{\hbox{eq}}} \approx 0.18 \, \mu$m, 
which is shorter than 
$r_{s1}= 0.43 \, \mu$m. At the same time, for the faster ones, 
$r_{s2}$ is very large,
i.e. they are 
predominantly screened by the first mechanism.
We, therefore, approximate the combined effect of the two mechanisms by 
using the following expression for the screening radius:
\begin{equation}
r_s^{-1}(\tau) = r_{s1}^{-1} + r_{s2}^{-1}(\tau).
\label{rs}
\end{equation}
The typical value of $r_s$
given by Eq.(\ref{rs}) is $r^{\ast}_s =0.2 \, \mu$m.

\section{evaluation of the noise spectrum}
\label{evaluation}

Now we are in position to evaluate
$\langle \phi(t, {\mathbf{r}})  \phi(0, {\mathbf{r}}^{\prime}) \rangle$
with $\phi(t, {\mathbf{r}})$ given by Eq.(\ref{phi}). 

Since an electron emitted by one deep
defect is most likely absorbed not by another defect but by the contact 
layer,  it is appropriate to assume that different defects 
fluctuate independently, which implies that, 
for $i \neq j$, $\langle \Delta q_i(t) \Delta q_j(0) \rangle = 0$.
Keeping also in mind that 
\begin{equation}
\langle \Delta q_i(t) \Delta q_i(0) \rangle =
\langle \Delta q_i^2 \rangle \exp (-t/\tau_i),
\label{qq}
\end{equation} 
we take the disorder average
of $\phi(t, {\mathbf{r}})  \phi(0, {\mathbf{r}}^{\prime})$, i.e.
we replace the discrete summation by the integrations
over the relevant probability distributions. This gives
\begin{widetext}
\begin{eqnarray}
\nonumber
\langle \phi(t, {\mathbf{r}})  \phi(0, {\mathbf{r}}^{\prime}) \rangle =
{n_D \over \epsilon^2} \int dE_B P(E_B) &
\left\{ \int  dE \, p(E)  \,
F({\mathbf{r}}, {\mathbf{r}}^{\prime}, \tau_+(E,E_B)) \, 
\langle \Delta q(E)^2 \rangle_+ \,
\exp \left(-{t \over \tau_+(E,E_B)} \right) \ \ \ \ \ \ \ \ \   \right. \\
& \left.  +
 \int  dE^{\prime} p^{\prime}(E^{\prime}) \, 
F({\mathbf{r}}, {\mathbf{r}}^{\prime}, \tau_-(E^{\prime},E_B)) \,
\langle \Delta q(E^{\prime})^2 \rangle_- \,
\exp \left( -{t \over \tau_-(E^{\prime},E_B)}\right) \! \right\},
\label{phiphi1}
\end{eqnarray}
\end{widetext}
where
\begin{eqnarray}
\nonumber
F({\mathbf{r}}, {\mathbf{r}}^{\prime}, \tau) && \equiv 
\int
{
 \hbox{exp} \left( - 
 {
  |{\mathbf{r}}- {\mathbf{a}}| + |{\mathbf{r}}^{\prime} - {\mathbf{a}}| 
 \over
  r_s(\tau)
 } 
 \right)
\over 
 |{\mathbf{r}}- {\mathbf{a}}| \ |{\mathbf{r}}^{\prime} - {\mathbf{a}}|
}
d^3 {\mathbf{a}} 
\\
&&
\approx 2 \pi r_s(\tau) 
\exp \left( - {|{\mathbf{r}}- {\mathbf{r}}^{\prime}|  
\over r_s(\tau) }  \right).
\label{F}
\end{eqnarray}
The approximation in Eq.(\ref{F}) can be justified by observing that the main 
spatial dependence of the integral for 
$F({\mathbf{r}}, {\mathbf{r}}^{\prime}, \tau)$ 
has form
$\exp \left( - {|{\mathbf{r}}- {\mathbf{r}}^{\prime}|
\over r_s(\tau) }  \right)$, and then the natural choice for
the prefactor is $F(0, 0, \tau) = 2 \pi r_s(\tau)$
(obtained by direct integration).

From Eq.(\ref{phiphi1}), the value of $|e \phi(t, {\mathbf{r}})|$ can 
be estimated as:
$e \sqrt{\langle \phi^2(0, 0) \rangle} \approx e^2   \epsilon^{-1} 
\sqrt{2 \pi n_D \nu(T) k_B T r^{\ast}_s } \sim 3.5\hbox{meV}$.
Since $k_B T \sim 30$meV, 
the  assumption $|e \phi(t, {\mathbf{r}})| \ll k_B T$ made earlier
was adequate.

Finally, we evaluate $C_V(t)$ by substituting Eq.(\ref{phiphi1}) into
Eq.(\ref{Cvphi}), and then take the Fourier transform (\ref{Sv}) to
obtain 
\begin{eqnarray}
\nonumber
{S_V(f) \over V^2} = 
{ 2 \pi^2 e^4 n_D \over \epsilon^2 (k_B T)^2 A z_T^2}
\ \ \ \ \ \ \ \ \ \ \ \ \ \ \ \ \ \ \ \ \ \ \ \ \ \ \ \ \ \ \ \ \ \ 
\\
\nonumber
\times  \int dE_B P(E_B) 
\left\{ \int  D(f, \tau_+(E, E_B)) \tilde{p}(E) dE 
\right. 
\ \ \ \ \ 
\\
+ 
\left. \int  D(f, \tau_-(E^{\prime}, E_B)) 
\tilde{p}^{\prime}(E^{\prime}) dE^{\prime}  
  \right\},
\label{Sv1}
\end{eqnarray}
where 
\begin{equation}
 D(f, \tau) = \!
{\tau r_s^4(\tau) \! \left[4 z_T - 3 r_s(\tau) + 
\hbox{\textsl{e}}^{ -{2 z_T \over r_s(\tau)} }
(2 z_T + 3 r_s(\tau)) \! \right] 
\over 
1 + 4 \pi^2 f^2 \tau^2} .
\label{D}
\end{equation}

Recalling the approximations
associated with (i) the assumption of the constant resistivity layer;
(ii) the assumption of the constant profile of defect concentration; 
(iii) the treatment of the screening
radius; and (iv) the evaluation of 
$F({\mathbf{r}}, {\mathbf{r}}^{\prime}, \tau)$ (Eq.(\ref{F})),  we
estimate that the  integrated noise intensity obtained from 
Eq.(\ref{Sv1}) entails a factor of two theoretical uncertainty
for the noise mechanism considered.

The complex appearance of Eq.(\ref{Sv1}) is due to the fact that it includes
separate terms for the two transitions \mbox{$D^+ \leftrightarrow D^0$} 
and \mbox{$D^0 \leftrightarrow D^-$}. 
Such a separation is necessary only
because the expressions (\ref{tau+0}, \ref{tau0-})
for the activation times $\tau_+$ and $\tau_-$ are slightly different from each other. 
This difference, however, vanishes 
if one assumes that the  width  of the distributions $\tilde{p}(E)$ and
$\tilde{p}^{\prime}(E^{\prime})$ is much smaller than the width of  $P(E_B)$,
i.e. $2 k_B T \ll \Delta E_B$. 

The complicated form of the numerator in the right-hand side of
Eq.(\ref{D}) is the analytic result of the
two integrations (\ref{Cvphi})  
over the volume of the ``constant resistivity'' layer.   
This expression reflects 
the competition between
two length scales: the  half-thickness  of that 
layer ($z_T$) and the screening radius ($r_s$). In the ``thin layer limit''
$z_T \ll r_s$,  the expression in the square brackets in Eq.(\ref{D}) 
can be approximated by ${2 z_T^2 \over r_s}$, whereas in the
``bulk limit'' $z_T \gg r_s$ that expression approaches
$4 z_T$. 

Now we simplify Eq.(\ref{Sv1}) by taking limits
$z_T \ll r_s$, $2 k_B T \ll \Delta E_B$ and
substituting $\nu(T) = {1 \over 2 \Delta E}$ 
(just an approximate numerical fact),
which gives 
\begin{equation} 
{S_V(f) \over V^2} = 
{ 8 \pi^2 e^4 n_D \over \epsilon^2 k_B T \Delta E A }
\int {\tau_0(E_B) 
r_s^3 \left( \tau_0(E_B) \right) P(E_B) dE_B \over 1 + 4 \pi^2 f^2 \tau_0^2(E_B)},
\label{Sv2}
\end{equation}
where 
\begin{eqnarray}
\nonumber
{1 \over \tau_0(E_B)} & \equiv & {1 \over \tau_+(\mu, E_B))} 
\equiv {1 \over \tau_-(\mu, E_B))} \\
& = & 1.5 \ \omega_0(E_B) \exp \left(-{E_B - \mu \over k_B T}\right).
\label{tauEB}
\end{eqnarray}
Our actual system is characterized only by the weaker inequalities
$z_T < r_s$, $2 k_B T < \Delta E_B$. Therefore, for comparison with
experiments we shall still use the original 
formula (\ref{Sv1}).  At the same time, for the qualitative analysis,
we will focus on Eq.(\ref{Sv2}), 
but all the conclusions will be fully applicable 
to Eq.(\ref{Sv1}).

A remarkable feature of Eq.(\ref{Sv2}), which can be traced back 
to Eq.(\ref{Cvphi}), is that, even though the resistance noise is 
caused by the fluctuations in 
the number of conduction electrons,
the resulting noise spectrum 
is independent of their equilibrium concentration.  Furthermore, that
spectrum is  
only weakly dependent on all the defect parameters involved 
(see Ref.~\cite{numbers}). 
In particular, the  dependence of the prefactor on the concentration
of defects $n_D$ is balanced by
the $n_D^{-1/2}$ dependence of $r_{s2}$, which then enters $r_s^3$ via 
Eq.(\ref{rs}). 

Like in the BDAE model, 
the $1/f$-like spectral shape generated by formula (\ref{Sv2})
is the result of the broad distribution of the activation energies $P(E_B)$,
but, at the same time, formula (\ref{Sv2})
has also two new features,
namely: (i)
the $1/T$ dependence of the prefactor and (ii) the energy-dependent weight
$r_s^3 \left( \tau(E_B) \right)$ multiplying $P(E_B)$. The second feature is 
particularly important for extracting the correct distribution
of $P(E_B)$ from experimental data.

For comparison with experiment in Section~\ref{experiment}
we will need the integral of ${S_V(f) \over V^2}$ over all frequencies, 
which is equal to $C_V(0)$. 
The expression for $C_V(0)$ can be obtained by substituting 
${1 \over 4} r_s^4(\tau) \! \left[4 z_T - 3 r_s(\tau) + 
\hbox{\textsl{e}}^{ -{2 z_T \over r_s(\tau)} }
(2 z_T + 3 r_s(\tau)) \! \right]$ instead of $D(f,\tau)$
into Eq.(\ref{Sv1}).
The mathematical structure of this expression
is similar to Eq.(\ref{Sv1}) but otherwise
not very illuminating to be written explicitly one more time. 
Instead, we give an estimate for the noise integral
corresponding to the thin film limit (\ref{Sv2}).
We do it with one further simplification. Namely,
we replace $r_s(\tau(E_B))$ in Eq.(\ref{Sv2})
by the typical value $r_s^{\ast} = 0.2 \ \mu$m  and then
obtain:
\begin{equation}
\int_0^{\infty} {S_V(f) \over V^2} \  df = 
{ 2 \pi^2 e^4 n_D  {r_s^{\ast}}^3 \over \epsilon^2 k_B T \ \Delta E \ A }.
\label{integral1}
\end{equation}

Although the bulk limit $z_T \gg r_s$ appears to be 
incompatible with the assumption of
independent defect fluctuations 
(to be explained in Section~\ref{generalizations}),
it is still instructive to give  an expression for
the noise integral in this case too. Combining the limit $z_T \gg r_s$
with all other 
approximations  used to derive Eq.(\ref{integral1}), we obtain
\begin{equation}
\int_0^{\infty} {S_V(f) \over V^2} \  df = 
{ 4 \pi^2 e^4 n_D  {r_s^{\ast}}^4 \over \epsilon^2 k_B T \ \Delta E \ A z_T }.
\label{integral2}
\end{equation}

Up to numerical prefactors, both approximations (\ref{integral1}) and 
(\ref{integral2}) can be summarized  as follows:
the noise intensity is proportional to $\left({e \over k_B T}\right)^2$, 
multiplied by the
mean squared amplitude of potential fluctuations 
${e^2 n_D k_B T r_s^{\ast} \over \epsilon^2 \Delta E}$,
further multiplied by the  
screening volume of a typical fluctuation ${r_s^{\ast}}^3$,
and, finally, divided by the volume of the space, 
where the defects contributing 
to the potential fluctuations are located. In the thin layer limit
that volume is roughly $2 r_s A$, whereas in the bulk limit
it is $2 z_T A$. The combination $n_D k_B T/\Delta E$  appearing in
the estimate of the potential fluctuation
should be identified with the concentration of ``thermally active'' defects, i.e.
those defects that fall in the thermal energy window 
around the chemical potential.

Now we give various estimates of the prefactor
in front of the approximate $1/f$ dependence of the spectrum.
For the thin film limit, we start from Eq.(\ref{Sv2}) and then, assuming 
$r_s(\tau(E_B)) = r_s^{\ast}$, $P(E_B) = {1 \over 2 \Delta E_B}$
and $\omega_0(E_B) = \omega_0(E_{B0})$, obtain 
\begin{equation} 
{S_V(f) \over V^2} = 
{\pi^2 e^4 n_D  {r_s^*}^3 \over \epsilon^2 A \ \Delta E \ \Delta E_B} 
\ {1 \over f}.
\label{Sv3}
\end{equation}
For the bulk limit, the analogous expression is
\begin{equation} 
{S_V(f) \over V^2} = 
{ 2 \pi^2 e^4 n_D  {r_s^*}^4 \over 
\epsilon^2 A z_T \ \Delta E \ \Delta E_B} \ {1 \over f}.
\label{Sv4}
\end{equation}

Equation(\ref{Sv4}) admits yet another remarkable simplification, 
if one substitutes 
$r_{\hbox{eq}}$ given by Eq.(\ref{rseq}) instead of $r_s^*$. 
(The  equilibrium self-screening mechanism described by Eq.(\ref{rseq}) 
should indeed be 
a proper description for the slower fraction of the defects, 
which means the lower 
frequency part of the spectrum.) In this case
\begin{equation} 
{S_V(f) \over V^2} = 
{ \Delta E \over 2 A z_T n_D \Delta E_B} \ {1 \over f} \sim 
{ 1 \over N_D f},
\label{Sv5}
\end{equation}
where $N_D = 2 A z_T n_D$ is the total number of defects in the
constant resistivity layer.
[In the above approximation, we assumed $\Delta E \sim \Delta E_B$.]
Although this result looks like a standard statistical factor,
its origin is, in fact, far more complex. It can be traced back to 
an accidental
interplay between the resistivity fluctuations and 
the screening mechanism.  Given the limitations
of the ``bulk limit'', it is unlikely that formula (\ref{Sv5})
represents a clean limit in any realistic system. At the same time,
for $r_s \sim z_T$, this formula still gives a reasonable estimate.

It is, finally, interesting to observe that, 
for the numbers characterizing our system, the Hooge's formula\cite{Hooge} 
${S_V(f) \over V^2} =  {\alpha \over N_C f}$ with coefficient 
$\alpha \sim 10^{-3}$,
would produce an estimate close to that of Eq.(\ref{Sv5}), (Here $N_C$ 
is the total number of conduction electrons.)
This simply reflects the fact, that in our case the concentration
of conduction electrons is about thousand times smaller than the 
concentration of defects.
The exact ratio of these concentrations is strongly temperature dependent,
as is the value of $\alpha$ required to fit 
the noise intensity in our experiments.

\section{Comparison with experiment}
\label{experiment}

\begin{figure}
\setlength{\unitlength}{0.1cm}
\begin{picture}(50, 74)
{
\put(-40, -7){
\epsfxsize= 5.3in
\epsfbox{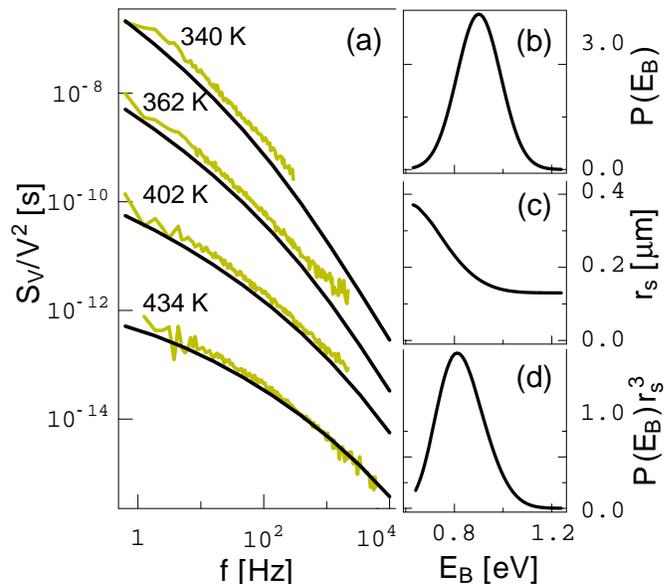} 
 }
}

\end{picture} 
\caption{(a) Noise spectra: 
erratic lines --- experiment, smooth lines ---  
predictions of Eq.(\protect{\ref{Sv1}}). 
The spectra at 340K, 362K and 402K are multiplied, respectively, by 
$30^3$, $30^2$ and $30$ to make them distinguishable.
On the real scale all four spectra fall almost on the top of each other. 
(b-d) Probability distribution $P(E_B)$, 
the screening radius $r_s$ and the product 
$P(E_B) r_s^3$ as functions of the barrier energy $E_B$ counted from
$\mu=0$.
} 
\label{fig2} 
\end{figure}

Theoretical spectra computed from Eq.(\ref{Sv1}) 
are compared with the experimental ones 
in Fig.~\ref{fig2}a (see also Ref.~\cite{supplement}).
The experimental spectra  were 
obtained by subtracting the zero-current noise from the total noise observed
with $V=50$~meV.
They were consistent with the spectra reported in Ref.~\cite{Verleg}, 
though the film was newly grown. 
All the theoretical spectra were obtained with 
$P(E_B)$ shown in Fig.~\ref{fig2}b
and characterized by \mbox{$E_{B0} = \mu + 0.90 \,$eV}
and $\Delta E_B = 0.09\,$eV. The uncertainties of the  fit  for
$E_{B0}$  and $\Delta E_B$ are $0.05$~eV and $0.02$~eV, 
respectively. 
Figure~\ref{fig2}c shows  $r_s(E_B)$  obtained by averaging 
$r_s \left( \tau_+(E, E_B)\right)$ and 
$r_s \left( \tau_-(E^{\prime}, E_B)\right)$ over $E$ and $E^{\prime}$ at
$T= 434\,$K.
Finally, Fig.~\ref{fig2}d 
illustrates the role of 
the $r_s^3$ weight by showing $P(E_B) r_s^3(E_B)$. 
If
$P(E_B)$ were to be obtained by fitting the experimental spectra to the
BDAE model \cite{Verleg}, the result would look like
Fig.~\ref{fig2}d, 
i.e. the maximum would be shifted by about 0.1~eV in
comparison with Fig.~\ref{fig2}b.

With the above value of $\Delta E_B$, the estimates 
(\ref{Sv3},\ref{Sv4},\ref{Sv5}) for the prefactor in front
of $1/f$ would give respectively:
$9 \cdot 10^{-12}$, $3 \cdot 10^{-11}$ and $1 \cdot 10^{-11}$ ---
all in reasonable agreement with experiments. [For the estimates
(\ref{Sv4}) and (\ref{Sv5}) we used $z_T = 0.13 \ \mu\hbox{m}$.]

\begin{figure}
\setlength{\unitlength}{0.1cm}
\begin{picture}(50, 50)
{
\put(-50, -180){
\epsfxsize= 7in
\epsfbox{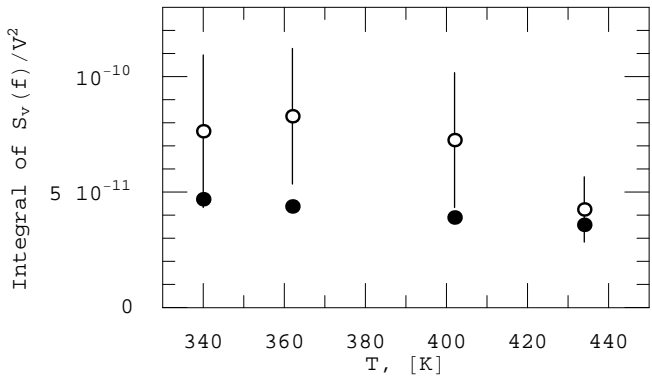} 
 }
}

\end{picture} 
\caption{Integrated noise intensity for the four spectra presented in
Fig.~\ref{fig2}. Empty dots
represent experimental values, and the solid black dots
theoretical values obtained from Eq.(\ref{Sv1}). The error bars 
on the experimental points are obtained as described in the text 
} 
\label{fig3} 
\end{figure}

In Fig.~\ref{fig3} we present another test of our theory, 
which is independent of the choice of $E_{B0}$ and $\Delta E_B$. Namely,
we compare the theoretical and experimental 
values of the integrated noise intensities for the four
temperatures indicated in Fig.~\ref{fig2}a. 

The experimental evaluation of  the integrated
intensity of an $1/f$-like noise is a
task notorious for its ambiguity. 
Fortunately, in our case, the power law extrapolations  of 
all four noise spectra were convergent,
which allowed us to use the following procedure:

First, we obtained the lower ends of the error bars by integrating the
experimental noise
spectra only in the frequency range of the actual experimental observations.
Then, the upper ends  were obtained
by making power law extrapolations of the spectra 
beyond the frequency range of observation 
(up to $10^{-6}$~Hz for small frequencies and $10^8$~Hz for large
frequencies), and then
adding the integrals over the extrapolated tails to the lower end values
of the error bars.
Finally, the ``experimental''
points indicated in Fig.~\ref{fig3} were  chosen as
the middle points of the above error bars.

The theoretical points presented in  Fig.~\ref{fig3}
were obtained for the full spectrum (\ref{Sv1}) as described in 
Section~\ref{evaluation}. 
One can also check that the  estimate (\ref{integral1})  gives  
numbers, which are not much different, e.g.:
$5.6 \cdot 10^{-11}$  for $T= 340$~K,
 and  $4.4 \cdot 10^{-11}$ for  $T= 434$~K.  

Given the large uncertainties of the extrapolation and
the theoretical uncertainties indicated earlier, 
the factor of two agreement between the 
theoretical and experimental points in Fig.~\ref{fig3}
should be considered as a successful consistency 
test of the noise mechanism proposed.

\section{Possible generalizations}
\label{generalizations}

Now we discuss briefly various modifications
required in order to generalize our treatment
to $1/f$ noise experiments  in other settings involving a-Si:H.

The application of the present 
theory to the  films with coplanar 
currents\cite{KK,GJK,Johanson,BK,Kasap} and also to the bulk samples
would encounter an 
essential complication related to the absence of contact layers.
In this case, the charge fluctuations of different defects become correlated
via emission and capture of the same electron. In such a process, the potential 
fluctuation is simply translated in space from one defect to another,
which means that, unless the two defects have different screening radii,
the total resistance does not fluctuate at all. Our analysis in 
Section~{\ref{defects}} revealed one possibility for different
defects to have different screening radii 
(as a function of their activation times). However, even with
this possibility, the overall effect of the absence
of contact layers should be a noticeable reduction of the noise intensity.

The contact layers also play a role in the screening of defects.
Without them (and without the $n$-layers) the first screening mechanism 
considered in Section~{\ref{defects}}
is not operational. Therefore, 
the treatment of screening of the faster fluctuating
defects becomes more difficult. 

The theory has to be further modified to include 
the hole conduction,
if, as a result of doping, the chemical potential
moves in the middle or below the middle of the band gap. 
An interesting situation 
may arise, when hole resistivity is equal to electron resistivity.
In this case one should expect a drastic reduction of the noise intensity, 
because the change of electron resistivity induced by potential fluctuations 
will be compensated by the change in the hole resistivity.

Analysing experiments,  
one should also keep in mind that  creation and  annealing of
charged defects represents an alternative way to 
potential fluctuations.
In particular, in a-Si:H such a process may be associated with 
the diffusion of hydrogen atoms\cite{Street}, which can inhibit or expose the
dangling bonds. The parameters describing this process are
not very well known. In our case, the mechanism involving the 
emission and the capture of conduction electrons by already 
existing defects appears
to be sufficient, but, in general, 
creation and annealing of defects can lead 
to a comparable  contribution to the integrated noise intensity.

Summarizing this Section, we would like to emphasise that 
all the above complications can still be treated on the basis of 
formula~(\ref{Cvphi}), but different ingredients for the evaluation 
of the potential fluctuations will be required.

\section{Conclusions}

In conclusion, we have developed a microscopic theory of 
$1/f$ noise in $n-i-n$ sandwich structures of a-Si:H  
and found  
a very good agreement between this theory and our experiments. 
The noise mechanism proposed 
should be quite general (in particular, in semiconductors): 
it merely requires the presence of charge traps
in regions of poor screening. The full calculation
of the actual noise spectrum necessarily involves  
a fair amount of material specific details, 
which, to some extent, mask the generality of our treatment.
However, one should remember,  that our calculation is
based on formula (\ref{Cvphi}), which
appears to have quite a broad range of applicability.

From a different perspective,
our analysis introduces a new method of characterizing the deep defects
in \mbox{a-Si:H}. 
In particular, our finding, that \mbox{$E_{B0}-E_{c0} = 0.27 \,$eV,}
which is substantially greater than 0.1~eV 
--- the scale of potential
fluctuations seen in the drift mobility experiments \cite{Street}, 
may indicate that the defects are not distributed randomly in
the background of the medium-range random potential but, instead, 
located in the regions of greater local strain, i.e. around the 
peaks of that potential.
 
Our result also limits the spectral intensities due to all other 
possible noise mechanisms to the difference between the experimental spectra
and the spectra given by Eq.(\ref{Sv1}). Even with the
freedom of varying $E_{B0}$  and $\Delta E_B$, this is
a very strong constraint.

\acknowledgements

We thank R. E. I. Schropp for providing us with the sample, and
A.~Buchleitner, B.~Farid and M. Weissman
for their helpful comments on the manuscript. 
The work of B. V. F. was supported by the 
Foundation of Fundamental Research on Matter (FOM), which is sponsored by
the Netherlands Organization for the Advancement of Pure Research (NWO).


\begin{references}
\bibitem{numbers}
In what follows, the values of $n_D$, $E_0$ and $\Delta E$ were obtained
from numerical fit to the non-linear IV-characteristics of our film 
taken at high applied voltages (cf. \cite{BHD}).
The values of $U$, $m_e^*$, $\sigma$, $N_c(E_{c0})$ and $\varepsilon_c$ 
represent the most general practice (see e.g. \cite{Street}).
\end{references}
\end{document}